\documentclass[aps,prc,superscriptaddress,twoside,twocolumn,nofootinbib,%
showpacs,dvips]{revtex4}
\usepackage{mathrsfs,bm}
\usepackage{longtable,lscape}
\usepackage{txfonts}
\usepackage{amssymb}
\usepackage{indentfirst}
\usepackage{graphicx,,booktabs}
\usepackage{color}
\usepackage{amssymb}

\begin{document}

\title{The $\Sigma$(1385) photoproduciton  from proton within a Regge-plus-resonance
approach}

\author{Jun He}
\email{junhe@impcas.ac.cn}
\affiliation{
Theoretical Physics Division,Institute of Modern Physics,Chinese Academy of Sciences,Lanzhou 730000, China
}
\affiliation{
Research Center for Hadron and CSR Physics, Lanzhou University and Institute of Modern Physics of CAS, Lanzhou 730000, China
}
\affiliation{
State Key Laboratory of Theoretical Physics, Institute of
Theoretical Physics, Chinese Academy of Sciences, Beijing  100190,China
}
\date{\today}

\begin{abstract}

The interaction mechanism of the $\Sigma$(1385) photoproduction from
proton $\gamma p\to K^+\Sigma^0(1385)$ is investigated within a
Regge-plus-resonance approach based on the experimental data released
by the CLAS Collaboration recently. The $t$ channel and the $u$ channel
are responsible to the behaviors of differential
cross sections at forward and backward angles, respectively. The
contributions from nucleon resonances including $N^*$ and $\Delta^*$,
which are determined by the predicted decay amplitudes in the constituent
quark model, are found small, but the $F_{35}$
state $\Delta(2000)$ is essential to reproduce
differential cross section.

\end{abstract}

 \maketitle
\maketitle
\section{Introduction}

The study of nucleon resonances is an important topic of hadron
physics. The information about nucleon resonance is mainly
extracted from the pion-nucleon scattering especially in the early
stage of the study of nucleon resonance~\cite{Vrana:1999nt}.
Based on a large number of nucleon resonances found in the
experiment, the constituent quark model (CQM) was developed and
achieved great success in the explanation about the property of
nucleon resonance~\cite{Isgur:1978xj,Capstick:1986bm}. However, the
predicted nucleon resonances in the CQM are much more than the ones found
in the experiment, which is the so-called ``missing resonance''
problem. One explanation about this problem is that the decay ratio of ``missing
resonance'' is very small in the usual experimental detected
channels, such as the pion-nucleon channel. Hence, the channels more than
the pion-nucleon channel, such as $\eta N$ and strange channels, attract
much attentions.

A mount of experimental data of the kaon photoproduction companied by a
ground strange baryon $\Lambda$ or $\Sigma$ have been accumulated in
the recent years~\cite{Klempt:2009pi}. However, the study about the
kaon photoproduction with a strange baryon resonance is scarce.  Very
recently, the CLAS Collaboration released their experimental data about
the kaon photoproduction with $\Sigma$(1385), $\Lambda(1520)$ and
$\Lambda(1405)$ with high precision~\cite{Moriya:2013hwg}, which
provide an opportunity to study nucleon resonances in these
channels.

The strong decays of nucleon resonances to  $\Sigma$(1385),
$\Lambda(1520)$ or $\Lambda(1405)$ with kaon meson have been studied
in the CQM~\cite{Capstick:1998uh}.  Combined with the theoretical
prediction about the radiative decay~\cite{Capstick:1992uc}, one can
make a rough estimation which nucleon resonances play important roles
in a certain photoproduction process.  For example, the large decay
widths to $N\gamma$ and $K\Lambda(1520)$ suggest that $N(2120)$ should be
easy to be found in the kaon photoproduction with $\Lambda(1520)$,
which has been confirmed by many theoretical analysis of the
$\Lambda(1520)$ photoproduction data~\cite{Xie:2010yk,He:2012ud}.  The
CQM prediction suggests a large decay ratio of the $D_{13}$ state
$N(2095)$ and the $F_{35}$ state $\Delta$(2000) in $\Sigma(1385)K$ decay
channel\cite{Capstick:1998uh}.  Hence it is interesting to study the
roles played by such states in the $\Sigma$(1385) photoproduction.

There only exist some old experimental data about the $\Sigma$(1385)
photoproduction with low precision, which were obtained before last
seventies~\cite{CambridgeBubbleChamberGroup:1967zzb,Erbe:1970cq}.  The
LEPS Collaboration also released some results in this channel but only
at extreme forward angles~\cite{Niiyama:2008rt}. Correspondingly, the
theoretical studies are also scarce. In Ref.~\cite{Oh:2007jd}
the $\Sigma$(1385) photoproduction has been studied in an effective
Lagrangian approach based on the preliminary data from CLAS
Collaboration.  However, due to the absence of the constraint of the
precise data large discrepancies at low energies between the
experimental data and the theoretical predictions can be found in
the differential cross section released by the CLAS
Collaboration~\cite{Moriya:2013hwg}.  In this work, we will analyze
the new CLAS data within a Regge-plus-resonance approach and
investigate the roles played by nucleon resonances.

This paper is organized as follows. After introduction, we will
present the effective Lagrangian used in this work and Reggeized
treatment for $t$ channel. The gauge invariance will be also discussed
in this section. The numerical results for the  cross section will be
given and compared with the experimental data in Section~\ref{Sec:
Results}. Finally, the paper ends with a brief summary.

\section{FORMALISM}

\begin{figure}[ht!]
\includegraphics[bb=135 555 510 720, scale=0.73,clip]{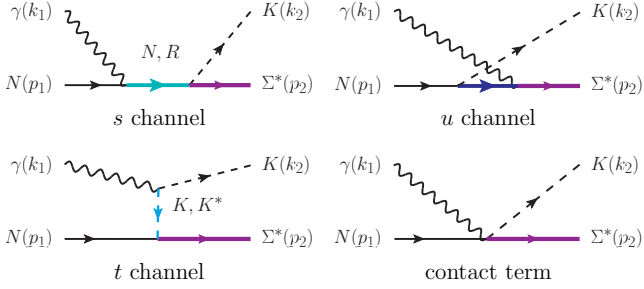}
\caption{(Color online) The diagrams for the $s$, $u$ and $t$ channels
and contact term for $\gamma p\to K^+\Sigma^0(1385)$.}
\label{pic:dia}
\end{figure}

The four types of interaction mechanism,  the $s$ channel,  the $u$ channel,
the $t$ channel and the contact term
for the $\Lambda(1520)$ photoproduction from nucleon with
$K$ are presented in Fig.~\ref{pic:dia}.
The Born terms contain the $N$, $Y$, $K$
intermediate states and the contact term.

For the Born $s$ channel, $t$ channel and contact term, the Lagrangians involved
are given as below,
\begin{eqnarray}
\mathcal{L}_{\gamma KK} &=& ie A_\mu \left( K^- \partial^\mu K^+ -
\partial^\mu K^- K^+
\right),\\
\mathcal{L}_{KN\Sigma^*} &=& \frac{f_{KN\Sigma^*}}{m_K} \partial_\mu
\overline{K} \overline{\bm \Sigma}^{*\mu} \cdot \bm{\tau} N + \mbox{
H.c.}, \\
\mathcal{L}_{\gamma NN} &=& -e \bar{N} \left( e_N\gamma^\mu
- \frac{\kappa_N^{}}{2M_N} \sigma^{\mu\nu}
\partial_\nu\right) A_\mu N, \\
\mathcal{L}_{\gamma KN\Sigma^*} &=& -ie \frac{f_{KN\Sigma^*}}{m_K}
A^\mu K^- \left( \bar{\Sigma}_\mu^{*0} p + \sqrt{2}
\bar{\Sigma}_\mu^{*+} n \right)
+ \mbox{ H.c.},
\label{eq:lag1}
\end{eqnarray}
where $A^\mu$, $N$, $K$, $\Sigma^{*\mu}$ are the photon, nucleon, kaon and
$\Sigma^*(1385)$ fields and the charge of nucleon $e_N=1,0$ for proton and neutron in the unit of $e=\sqrt{4\pi \alpha}$ with $\alpha$ being the fine-structure constant . The anomalous
magnetic moment $\kappa_N=1.79$ for proton.  $m_K$ and $M_N$ are the
masses of kaon and nucleon.
The coupling constant for $KN\Sigma^*$ vertex can be related to the
$\pi N \Delta$ coupling by the SU(3) flavor symmetry relation,
and the value $f_{KN\Sigma^*} = -3.22$ can be obtained~\cite{Oh:2007jd,Gao:2010hy}.

The $t$ channel for the $\Sigma$(1385) photoproduction occurs through both
$K$ and $K^*$ exchanges. As shown in
Ref.~\cite{Oh:2007jd}, the contribution from the
$K^*$ exchange is negligible  at energies
$E_\gamma=3\sim4$ GeV with the reasonable coupling constant. Hence only
$K$ exchange will be considered in this work.

The $u$ channel
diagram shown in Fig.~\ref{pic:dia} contains intermediate
hyperons. The effective Lagrangians
for these diagrams are
\begin{eqnarray}
\mathcal{L}_{\Sigma^*Y\gamma} &=& -\frac{ief_1}{2M_Y} \overline{Y} \gamma_\nu
\gamma_5 F^{\mu\nu} \Sigma_\mu^*
- \frac{ef_2}{(2M_Y)^2} \partial_\nu \overline{Y} \gamma_5 F^{\mu\nu}
\Sigma_\mu^* + \mbox{ H.c.},\nonumber\\
\mathcal{L}_{KNY} &=& \frac{g_{KNY}^{}}{M_N +  M_Y} \overline{N} \gamma^\mu
\gamma_5^{} Y \partial_\mu K + \mbox{ H.c.},
\end{eqnarray}
where $F_{\mu\nu} = \partial_\mu A_\nu - \partial_\nu A_\mu$ and $Y$
stands for a hyperon with spin-$1/2$ carrying a mass $M_Y$ .
For the intermediate $\Lambda(1116)$ state,
$f_1 = 4.52$ and $f_2 = 5.63$~\cite{Oh:2007jd}.
The coupling constant $g_{KN\Lambda}$  can be determined by
flavor SU(3) symmetry relation, which gives value $g_{KN\Lambda}=
-13.24$~\cite{Oh:2007jd}. The $\Sigma$ exchange is negligible due to to the small
coupling constant determined from SU(3) symmetries~\cite{Oh:2007jd}.  Besides, the effect of
higher resonances can be included through Reggeized treatment.
Due to the large uncertainty at backward angles of the CLAS
experimental data and much larger masses of $\Lambda$ and $\Sigma$
baryons
compared with the mass of $K$ in the $t$ channel
we do not consider the Reggeized treatment in the $u$ channel in this work
as Ref.~\cite{Corthals:2005ce}.

The amplitude for the $u$ channel is gauge invariant itself while the
amplitudes for the Born $s$ channel, $t$ channel and contact term are not gauge
invariant. After summing up the amplitudes from the $s$ channel,
the $t$ channel and the contact term of the $\Sigma$(1385) photoproduction,
the gauge invariance can be guaranteed as the $\pi$ and $\Lambda(1520)$
photoproductions~\cite{Ohta:1989ji,Nam:2010au}. The effect
of the hadron internal structure can be reflected by the form factor
added at each vertex. Unfortunately, it will violate the gauge invariance. To
restore the gauge invariance, a generalized contact term is introduced
as~\cite{Haberzettl:2006bn}
\begin{eqnarray}
	{\cal M}^{\mu\nu}_c&=&\frac{ief_{KN\Sigma^*}}{m_K}\left[g^{\mu\nu}F_t+
		k_2^\mu
		(2k_2-k_1)^\nu\frac{(F_t-1)[1-h(1-F_s)]}{t-m_K^2}\right.\nonumber\\
		&+&\left.k_2^\mu
		(2p_1-k_1)^\nu\frac{(F_s-1)[1-h(1-F_t)]}{s-M_N^2}\right],
		\label{Eq: contact}
\end{eqnarray}
where the $h$ is a free parameter and will be fitted and $F_i$ with
$i=s,t$ is the form factor. 

In this work, for the Born $s$ channel and the $u$ channel we choose the form
factor in the from,
\begin{eqnarray}
F_i(q^2)&=&\left(\frac{n\Lambda_i^4}{n\Lambda_i^4+(q^2-M^2)^2}\right)^n,\label{Eq: FF}
\end{eqnarray}
which goes to Gaussian form as $n\to \infty$ and
for $t$ channel $K$ exchange,
\begin{eqnarray}
F_i(q^2)&=&\frac{\Lambda_i^2-M^2}{\Lambda_i^2-q^2},
\end{eqnarray}
where $M$ and $q$ are the mass and momentum of the off-shell
intermediate particle. The cutoff $\Lambda_i$ for $s$,
$u$ or $t$ channel should be about 1~GeV and will be
set as free parameter in this work.

We introduce $K$ Reggeized
treatment as following to describe the behavior of differential cross
section of the $\Sigma(1385)$ photoproduction at high photon energies
~\cite{Guidal:1997hy,Corthals:2005ce,Titov:2005kf},
\begin{eqnarray}
\label{eq:RT}
\frac{1}{t-m^{2}_{K}}\to\mathcal{D}_{K}
&=&\left(\frac{s}{s_{scale}} \right)^{\alpha_{K}}
\frac{\pi\alpha'_{K}}{\Gamma(1+\alpha_{K})\sin(\pi\alpha_{K})},
\end{eqnarray}
where $\alpha'_{K}$ is the slope of the trajectory and the scale factor
$s_{scale}$ is fixed at 1 GeV$^2$.
$\alpha_{K}$ is the linear trajectory of the $K$ meson, which is a function of $t$ assigned as
follows, $\alpha_{K}=0.70\,\mathrm{GeV}^{-2}(t-m^{2}_{K})$. The $K^*$
Reggeized treatment is analogous. There is no reason a priori that the
coupling constants for Reggeized treatment $f^{Reg}_{KN\Sigma^*}$ and
$f^{Reg}_{K^*N\Sigma^*}$  are same as those for the real $K$ and $K^*$ exchange~\cite{oai:arXiv.org:0711.3536}. The same observation applies to the Reggeized K* coupling. In this work we set
them as free parameters. We expect the difference should not be very
large, so the $K^*$ exchange is still very small and omitted in
this work as Ref.~\cite{Oh:2007jd}.

The Reggeized treatment should work completely at high photon energies
and interpolate smoothly to low energies. It is implemented by Toki
$et\ al.$~\cite{Toki:2007ab} and Nam and Kao~\cite{Nam:2010au} by
introducing a weighting function ${\cal R}$. Here we adopt the
treatment as,
\begin{eqnarray}
\label{eq:R}
&&\frac{F_t}{t-m^{2}_{K}}\to \frac{F_t}{t-m^{2}_{K}} {\cal R}=\mathcal{D}_{K}R+\frac{F_t}{t-m^{2}_{K}}(1-{R}),\,\,\,\,
\end{eqnarray}
where $R={R}_{s}{R}_{t}$ with
\begin{eqnarray}
\label{eq:RSRT}
{R}_{s}&=&
\frac{1}{2}
\left[1+\tanh\left(\frac{s-s_{\mathrm{Reg}}}{s_{0}} \right)
	\right],\nonumber\\
{R}_{t}&=&
1-\frac{1}{2}
\left[1+\tanh\left(\frac{|t|-t_{\mathrm{Reg}}}{t_{0}} \right) \right].
\end{eqnarray}
The free parameters $s_{Reg}$, $s_0$, $t_{Reg}$ and $t_0$ will be
fitted with the differential cross section.

As inclusion of the from factor $F_i$, the Reggeized treatment will
violate the gauge invariance and current conservation also. To restore the current conservation, we redefine the relevant amplitudes
,
\begin{eqnarray}
\label{eq:WT1}
i\mathcal{M}^{\mu\nu}&=&i\mathcal{M}^{\mu\nu}_{t}+i\mathcal{M}^{\mu\nu}_{s}+i\mathcal{M}^{\mu\nu}_{c}
\to
(i\mathcal{M}^{\mu\nu}_{t}+i\mathcal{M}^{\mu\nu}_{s}
+i\mathcal{M}^{\mu\nu}_{c})\mathcal{R}
\nonumber\\
&\equiv&i\mathcal{M}^{\mathrm{Regge}\  \mu\nu}_{t}+(i\mathcal{M}^{\mu\nu}_{s}
+i\mathcal{M}^{\mu\nu}_{c})\mathcal{R}
.\ \ \ \  \
\end{eqnarray}
With such definition, the relation $k_1^\mu\mathcal{M}^{\mu\nu}=0$ is satisfied.

For the nucleon resonance contributions, we adopt the Lagrangians for the
radiative decay,
\begin{eqnarray}
	\mathcal{L}_{\gamma N R(\frac{1}{2}^{\pm})} &=&\frac{e f_2}{2M_N}
	\bar{N} \Gamma^{(\mp)}\sigma_{\mu\nu}F^{\mu\nu} R \,+{\rm H.c.},\nonumber \\
\mathcal{L}_{\gamma N R(J^{\pm})} &=&\frac{-i^{n}f_1}{(2M_N)^{n}}
\bar{N}~\gamma_\nu \partial_{\mu_2}\cdots\partial_{\mu_{n}}
F_{\mu_1\nu}\Gamma^{\pm(-1)^{n+1}}R^{\mu_1\mu_2\cdots\mu_{n}}\nonumber\\
&+&\frac{-i^{n+1}f_2}{(2M_N)^{n+1}} \partial_{\nu}\bar{N}
~ \partial_{\mu_2}\cdots\partial_{\mu_{n}}
F_{\mu_1\nu}\Gamma^{\pm(-1)^{n+1}}R^{\mu_1\mu_2\cdots\mu_{n}}\nonumber\\
&+&{\rm H.c.},\label{Eq:Lg}
\end{eqnarray}
where $F^{\mu\nu}=\partial^\mu A\nu-\partial^\nu A_\mu$ with
$R_{\mu_1\cdots\mu_n}$ is the field for the
nucleon resonance with spin $J=n+1/2$, and
$\Gamma^{(\pm)}=(i\gamma_5,1)$
for the different resonance parity. The Lagrangians here are also adopted
from the previous works on nucleon resonances with
spins $3/2$ or $5/2$~\cite{Nam:2010au,Xie:2010yk,Oh:2007jd}.

The Lagrangians for the strong decay can be written as
\begin{eqnarray}
\mathcal{L}_{R(\frac{1}{2}^{\pm})K\Sigma^*} &=& \frac{h_2}{2m_{K}}
\partial_\mu K\bar{\Sigma}^*_{\mu} \Gamma^{(\pm)}R,+{\rm h.c.}, \nonumber\\
\mathcal{L}_{R(J^\pm)K\Sigma^*}
&=&\frac{-i^{n+1}h_1}{m_K^{n}} \bar{\Sigma}^*_{\mu_1}~\gamma_\nu\partial_\nu
\partial_{\mu_2}\cdots\partial_{\mu_{n}}
K\Gamma^{\pm(-1)^{n}}~R^{\mu_1\mu_2\cdots\mu_{n}}\nonumber\\
&+&\frac{-i^{n}h_2}{m_K^{n+1}} \bar{\Sigma}^*_{\alpha}~\partial_{\alpha}\partial_{\mu_1}
\partial_{\mu_2}\cdots\partial_{\mu_{n}}
K\Gamma^{\pm(-1)^{n}}~R^{\beta\mu_1\mu_2\cdots\mu_{n}}\nonumber\\
&+&{\rm H.c.}.\label{Eq:Ls}
\end{eqnarray}
In this work the coupling constants $f_1$, $f_2$, $h_1$ and $h_2$ will
be determined by the helicity amplitudes $A_{1/2}$ and $A_{3/2}$ and
the decay amplitudes $G(\ell_1)$ and $G(\ell_2)$, which are obtained
in the CQM. The interested reader is referred to Refs.~\cite{Oh:2007jd,He:2012ud} for further information.

In this work the nucleon resonances $R$ including $N^*$ and $\Delta^*$
will be considered. The
resonance field $R$ carries either isospin-$1/2$ or isospin-$3/2$. By
omitting the space-time indices, the isospin structure of
$RK\Sigma^*$ vertex reads as,
\begin{eqnarray}
\overline{R} \bm{\Sigma}^* \cdot \bm{\tau} K=p\Sigma^0{K}^+
-n{\Sigma}^0{K}^0 +\sqrt{2}n{\Sigma}^- K^+
+\sqrt{2}p\Sigma^+ K^0,\ \
\end{eqnarray}
for resonance $R$ with isospin-$1/2$. If the resonance $R$ carries
isospin-$3/2$, the effective Lagrangian has the isospin structure as
\begin{eqnarray}
\overline{R}  \bm{\Sigma}^*\cdot\bm{T}
K&=&\sqrt{3}\Delta^{++}\Sigma^+K^+-\sqrt{2}\Delta^+\Sigma^0 K^+
-\Delta^0\Sigma^- K^+\nonumber\\&-&\sqrt{2}\Delta^0\Sigma^0 K^0
+\Delta^+\Sigma^+ K^0-\sqrt{3}\Delta^-\Sigma^- K^0.\ \ \ \ \ \
\end{eqnarray}

\section{Results}\label{Sec: Results}

As shown in the previous section, we consider the $s$ channel with
intermediate nucleon, the Reggeized $t$ channel with $K$ exchange, the $u$ channel with
intermediate $\Lambda$, the contact term and the nucleon resonance
intermediate $s$ channel in the $\Sigma(1385)$
photoproduction.  By using the MINUIT code the differential cross section released by
the CLAS Collaboration recently will be fitted with the help of the Lagrangians
presented in the previous section. The free parameters involved and
their fitted
values are listed in Table~\ref{Tab: Para}. Here we exclude
total cross section in the fitting procedure because it can be
obtained by integrating the differential cross section.
\begin{table}[h!]
\renewcommand\tabcolsep{0.16cm}\renewcommand{\arraystretch}{1.5}
\begin{center}
\caption{The free parameters used in fittting. The cut-offs $\Lambda_i$ are in the
	units of GeV, the parameters  $s_{Reg}$, $s_0$, $t_{Reg}$ and $t_0$ for Reggeized treatment are in the
	units of GeV$^2$.\label{Tab: Para}}
\begin{tabular}{crcrcr}  \toprule[1pt]
$\Lambda_s$ &$0.77\pm0.03$ & $\Lambda_t$&$1.48\pm0.04$
&$\Lambda_u$&$0.98\pm0.01$
\\
$\Lambda_R$&$1.19\pm0.03$ & $h$&$1.66\pm0.07$\\\hline
$\sqrt{s_{Reg}}$&$2.10\pm0.01$  & $s_0$&$0.29\pm0.27$  &
$\sqrt{t_{Reg}}$&$4.95\pm3.62$    \\
 $t_0$&$0.88\pm0.48$  &$f^{Reg}_{KN\Sigma^*}$&$-4.74\pm0.02$    \\\bottomrule[1pt]
\end{tabular}
\end{center}
\end{table}

As expected, the fitted values of cutoffs $\Lambda_i$ for the $s$ channel,
the $t$ channel, the $u$ channel and the nucleon resonance contributions are close
to 1~GeV. The $s_{Reg}$ is about 2.1 GeV, which indicte the Reggeized
treatment play an important role even at the energies not so high as
the $K$ photoprodcution with $\Lambda$ baryon in
Ref.~\cite{Corthals:2005ce}.  The coupling constant
$g^{Reg}_{KN\Sigma^*}$ for the Reggeized $t$ channel is about half
larger than the values obtained by SU(3) symmetry, which is consistent
with our expectation. As will be presented, the experimental CLAS data
are well reproduced with $\chi^2=0.8$ per degree of freedom. If the
systematic uncertainties are excluded, the best fitted $\chi^2$ per
degree of freedom is $2.5$. The results of the cross section are
similar to  those with systematic uncertainty.

\subsection{Contributions from nucleon resonances}

First, we will present the contributions from  nucleon resonances.
As predicted in the CQM, for the $\Sigma(1385)K$ channel the decay widthsof nucleon resonances, such
as $N(2095)$ and $\Delta(2000)$, are large and expected to play more
important roles than
other nucleon resonances~\cite{Capstick:1998uh,Capstick:1992uc}. In this work we use following criterion
to select the resonances which will be considered in the fitting,
\begin{eqnarray}
\lambda=(A_{1/2}^2+A_{3/2}^2)(G(\ell_1)^2+G(\ell_2)^2)I^2\cdot10^{5}>\lambda_0,
\end{eqnarray}
where helicity amplitudes $A_{1/2,3/2}$ and partial wave decay amplitudes
$G(\ell)$ are in the units of $10^{-3}/\sqrt{\rm{GeV}}$ and
$\sqrt{\rm{MeV}}$. A factor $10^5$ is introduced to make the
largest value of $\lambda$ in the order of $10^0$. The isospin factor $I=1$  for $N^*$ and $\sqrt{2}$ for
$\Delta^*$.
\begin{table*}[htp!]
\renewcommand\tabcolsep{0.454cm}
\renewcommand{\arraystretch}{1.5}
\caption{The nucleon resonances considered. The mass $m_R$, helicity
	amplitudes $A_{1/2,3/2}$ and partial wave decay amplitudes
	$G(\ell)$ are in the unit of MeV, $10^{-3}/\sqrt{\rm{GeV}}$ and
	$\sqrt{\rm{MeV}}$, respectively. The explanation about
	$\lambda$, $\delta\chi^2$ and $\delta\chi^2_r$ can be found in
text. In the full model $\chi^2=0.8~[2.5]$ per degree of freedom. The
values in bracket are obtained after excluding the systematic
uncertainties.}
\begin{tabular}{ll|rr|rr|rrr} \toprule[1pt]
 State & PDG&  $A^p_{1/2}$ &   $A^p_{3/2}$ &  $G(\ell_1)$ &
 $G(\ell_2)$  & $\lambda$ & $\delta\chi^2$ &$\delta\chi^2_r$\\\hline
 $[N\textstyle{3\over 2}^-]_3(1960)$ &$N(2120)D_{13}**$ & 36 &  $-43$  & $1.3 ^{+ 0.4}_{-
 0.4}$ & $1.4 ^{+ 1.3}_{- 1.3}$ &1.1&0.2~[0.8]&0.1~[0.4]\\
 $[N\textstyle{3\over 2}^-]_4(2055)$ & & 16 & 0 & $-2.5 ^{+
 1.0}_{-1.0}$ & $-2.5$ $^{+ 2.3}_{- 1.9}$ & 0.3&0.0~[0.0]&0.0~[0.0]\\
 $[N\textstyle{3\over 2}^-]_5(2095)$ & & $-9$   & $-14$  &
 $7.7 ^{+1.2}_{-1.2}$ &$-0.8$ $^{+0.7}_{-1.0}$  &1.7 &0.4~[3.3] &0.1~[0.1]\\\hline
 $[N\textstyle{3\over 2}^+]_3(1910)$ & & $-21$& $-27$ & $-1.9 ^{+ 1.9}_{- 7.3}$ &
 0.0 $^{+ 0.0}_{- 0.4}$  & 0.4 &0.0~[0.5]&0.0~[0.1]\\
 $[N\textstyle{3\over 2}^+]_5(2030)$ & &$-9$   &15  & 2.2 $^{+ 1.0}_{- 1.9}$ & $-0.2 ^{+
 0.1}_{- 0.3}$  & 0.2 &0.0~[0.0] &0.0~[0.0]\\\hline
 $[N\textstyle{5\over 2}^+]_2(1980)$ & & $-11$  &$-6$ & $-3.6 ^{+ 2.5}_{- 3.0}$ & $-0.1 ^{+ 0.1}_{-
 0.3}$ &0.2 &0.2~[0.5]&0.1~[0.3]\\\hline
  $[\Delta\textstyle{3\over 2}^-]_2(2080)$ & $\Delta(1940)D_{33}**$&
  $-20$  &$-6$  & $-4.1 ^{+ 4.0}_{- 1.5}$ &
  $-0.5 ^{+ 0.5}_{- 2.2}$  &1.5 &0.1~[1.1]&0.0~[0.0]\\
$[\Delta\textstyle{3\over 2}^-]_3(2145)$ & & 0  &10 & 5.2 $\pm $ 0.4 &
$-1.9 ^{+ 1.2}_{- 4.0}$ & 0.6 &0.2~[1.0]&0.0~[0.0]\\\hline
  $[\Delta\textstyle{5\over 2}^+]_2(1990)$ &$\Delta(2000)F_{35}**$  &
  $-10$ &$-28$ & 4.0 $^{+ 4.5}_{-
  4.0}$ & $-0.1 ^{+ 0.1}_{- 0.4}$  &2.8 &1.5~[14.7] &0.7~[4.5]\\
\bottomrule[1pt]
\end{tabular}
\label{Tab: Resonances}
\end{table*}
First we choose several nucleon resonances in descending order of $\lambda$. If the contributions and influences of the nucleon resonances with small $\lambda$ is negligible, we stop here. If not, more resonances would be added.
 According to such criterion nine
resonances survived with $\lambda_0=0.1$ are listed in Table.~\ref{Tab: Resonances}.

For the masses of the nucleon
resonances, the values  suggested  in the Particle Data Group (PDG)~\cite{PDG} are
adopted and for the nucleon resonances not listed in PDG, the prediction by the
CQM will be adopted~\cite{Capstick:1998uh,Capstick:1992uc}. To prevent the
proliferation of the free parameters, the Breit-Wigner widths for all nucleon
resonances are set to 500~MeV, which is consistent to the widths for the
$\Delta(2000)$ and $\Delta(1940)$  obtained in the multichannel partial-wave
analysis~\cite{Shrestha:2012ep,Anisovich:2011fc}.
As shown in Table~\ref{Tab: Para}, the fitted value of cutoff for the nucleon
resonances $\Lambda_R=1.19$~GeV with Gaussian from
factor, that is, the form of from factor in Eq.~(\ref{Eq: FF}) with $n\to \infty$.

\begin{figure}[h!]
  \includegraphics[ bb=170 740 680 1040 ,scale=0.61,clip]{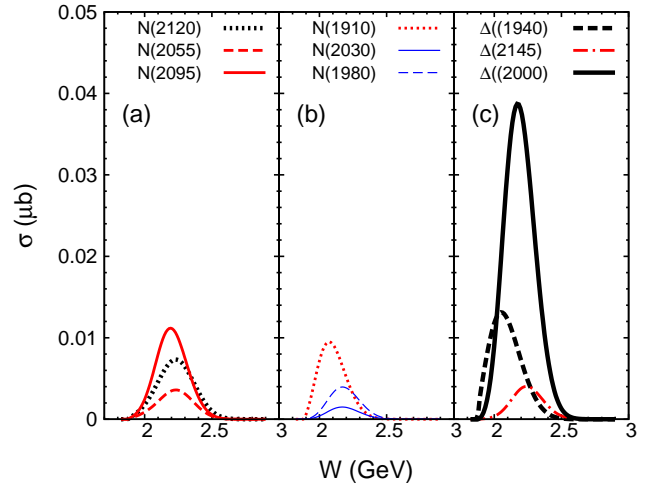}
  \caption{(Color online) Total cross section $\sigma$ for
	  corresponding
	  nucleon resonance as a function of the photon energy $W$ in center of mass frame.}\label{Fig: TCSR}
\end{figure}

In Fig.~\ref{Fig: TCSR}, we present the total cross section of each nucleon resonance
listed in Table~\ref{Tab: Resonances} to give an image of the magnitude of the
corresponding nucleon resonance. Generally, the contributions from the
nucleon resonances are smaller in the $\Sigma(1385)$ photoproduction compared
with the contributions of nucleon resonances in the $\Lambda(1520)$
photoproduction~\cite{He:2012ud}. The largest contribution is from $\Delta(2000)$ which
have largest $\lambda$ about 3. The three nucleon resonances listed in the PDG,
$N(2120)$, $\Delta(1940)$ and $\Delta(2000)$ have relatively large
contributions among all nucleon resonances considered. The
$N(2095)$ with largest decay width in
the $\Sigma(1385)K$ channel has much smaller contribution than
$\Delta(2000)$ due to its relative small radiatively decay width.

In Table~\ref{Tab: Resonances}, we list $\delta\chi^2$  and
$\delta\chi^2_r$, which are the
variations of the $\chi^2$ after turning off the corresponding
resonance without and with refitting, respectively. It reflects the
influence of the corresponding resonance on the reproduction of the
experimental
differential cross section. Generally, the variation of
the $\chi^2$ is consistent with the value of $\lambda$.  The
resonances with $\lambda>1$, $N(2120)$, $N(2095)$ and $\Delta(2000)$,
give $\delta\chi^2_r$ about or larger than 0.1~[0.7].
The $\Delta(2000)$ not only provides largest contribution to total
cross section as shown in Fig.~\ref{Fig: TCSR}, but also
has largest influence on the $\chi^2$ with
$\delta\chi^2_r=0.7~[4.5]$ after refitting. The influences of other
resonances including the $N(2095)$ are much smaller than $\Delta(2000)$. The $\lambda$ of
$N(2095)$ is large, about 1.7, while the $\delta\chi^2_r$ after refitting is about
0.1 which is much smaller than $\Delta(2000)$. It is due to the compensation effect from other resonances and (even) the Born terms in refitting. After a nucleon resonance turned off, the variation of the parameters after refitting will lead to the variation of the contributions from other resonances even the Born terms. The absence of the $N(2095)$ is smeared by such variation.

\subsection{The Contact term and the Reggeized treatment}

In this section we will present more explicit information about the
contact term and Reggeized treatment.
\begin{figure}[h!]
  \includegraphics[ bb=0 260 345 530 ,scale=0.6,clip]{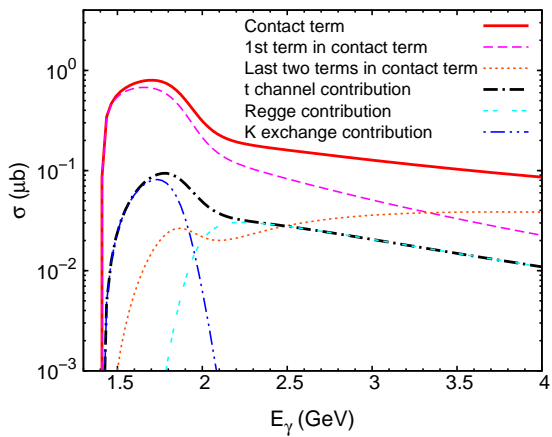}
  \caption{(Color online) Total cross section $\sigma$ as a function of the photon energy $E_\gamma$ for contact
  term and Reggeized treatment.}\label{Fig: tc}
\end{figure}
As shown in Fig.~\ref{Fig: tc}, the first term of the contact term in
Eq.~(\ref{Eq: contact}),
which comes from the Lagrangians given by Eq.~(\ref{eq:lag1}), play most
dominant role at energies upto about 3~GeV. For $t$ channel, the $K$
exchange is dominant in low energies while the Regge contribution
becomes dominant at energies higher than 2.5 GeV as we expected.

\subsection{Differential cross section}

\begin{figure}[h!]
  \includegraphics[ bb= 179 240 560 620 ,scale=0.58,clip]{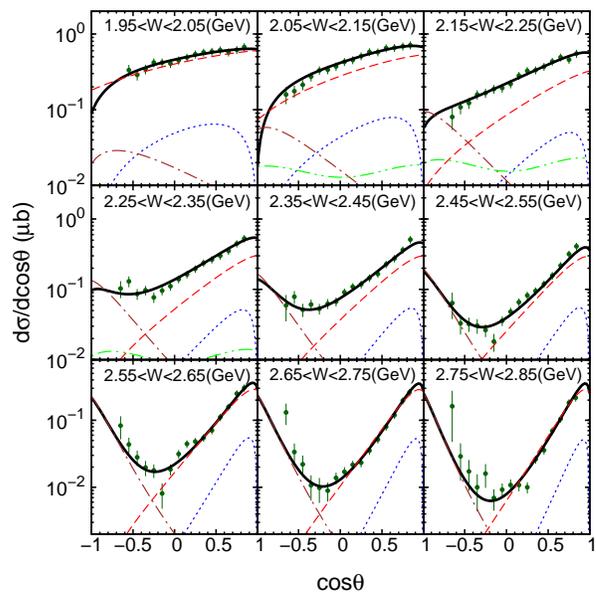}
  \caption{
(Color online) The differential cross section $d\sigma/d\cos\theta$ for
the $\Sigma$(1385) photoproduction from proton as a function of
$\cos\theta$. The full (black), dashed (red),
dash-dotted (brown), dotted (blue) and dash-dot-dotted (green) lines are for the full model,
the contact term, the  $u$ channel, the $t$ channel and $\Delta(2000)$, respectively. The data are from
\cite{Moriya:2013hwg}.}
	  \label{Fig: dcs}
\end{figure}

With the nucleon resonance contributions and the Born terms given in the previous
subsections,
the results of differential cross section for the $\Sigma$(1385)
photoproduction from proton compared with the CLAS data are shown in
Fig.~\ref{Fig: dcs}. As shown in the figure, the experimental data are well
reproduced in our model. The contributions from the $u$ channel and
the contact
term are dominant and is responsible for the behaviors of the
differential cross section at backward and forward angles,
respectively. The
$t$ channel contribution is smaller but give considerable
contribution at forward angles. The Born $s$ channel contribution is
very small.

Compared with the plausible results at forwards angles, the
results at backward angles are not so satisfactory.
 We have tried
to introduce the
Reggeized treatment $u$ channel contribution. But as mentioned in
Section II, the large uncertainty at backward angles make it difficult
to give an meaningful determination of extra five parameters
required by Reggeized treatment. Hence, we keep the $\Lambda$
intermediate $u$ channel in this work. The further experimental data at
extreme backward with high precision will be helpful to deepen the
understanding about the interaction mechanism in the $u$ channel.

\subsection{Total cross section}

\begin{figure}[h!]
  \includegraphics[ bb=0 259 345 530 ,scale=0.66,clip]{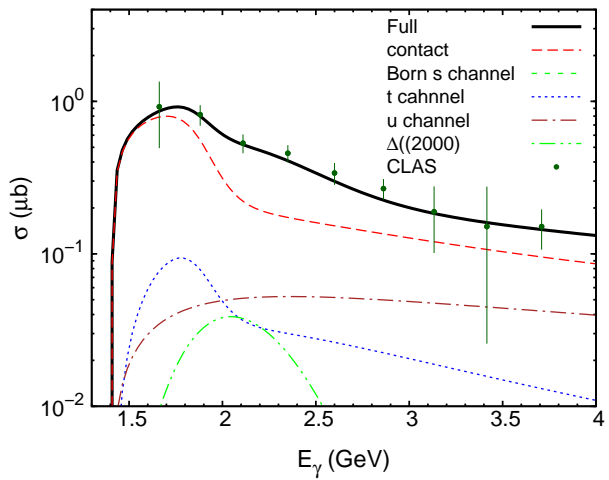}
  \caption{(Color online) Total cross section $\sigma$ as a function of the photon energy $E_\gamma$.
   The data are from
  Ref.~\cite{Moriya:2013hwg}.}\label{Fig: TCS}
\end{figure}

We also present the theoretical results of total cross section compared
with the CLAS data in Fig.~\ref{Fig: TCS}.
One can find that our result is well
comparable with the CLAS data. At all energies, the contact term
provides most important contribution, and the Reggeized $t$ channel
contribution is large near threshold and decreases rapidly at higher
energies. The $u$ channel contribution becomes important at higher
energies.  The contributions from the nucleon resonances are small. But
as shown in Table~\ref{Tab: Resonances}, it is essential to reproduce the differential cross
section. The $\Delta$(2000) has magnitude comparable to the $t$ channel
and the
$u$ channel at $E_\gamma$ about 2.1 GeV.

\section{Summary}

The $\Sigma$(1385) photoproduction in the
$\gamma p \to K^+\Sigma^0(1385)$ reaction is investigated within a
Regge-plus-resonance approach.  The contact term is dominant in the interaction mechanism and
Reggeized $t$ channel is important at energies near
threshold at forward angles. The $u$ channel is responsible for the behavior of
differential cross section at backward angles.

The contributions of nucleon resonances are determined by the
radiative and strong decay amplitudes predicted from the CQM. The results show that the
contributions from nucleon resonances are small compared with the
contact term, $u$ and $t$ channel contributions but essential to
reproduce the experimental data. The $D_{13}$ state $N(2095)$ which is expected to be
important in $\Sigma(1385)$ photoproduction have much smaller contribution for total
cross section and smaller influence on the reproduction of differential cross section than
$F_{35}$ state  $\Delta(2000)$. The resonance $\Delta(2000)$ is the most
important nucleon resonance in $\Sigma(1385)$ photoproduction as
suggested by CQM~\cite{Capstick:1998uh,Capstick:1992uc}, which is also
consistent with the results in Ref.~\cite{Oh:2007jd}.

\section*{Acknowledgement}

The author wants to thank Prof. Schumacher for valuable
comments on the manuscript.
This project is partially supported by the Major State
Basic Research Development Program in China (No. 2014CB845405),
the National Natural Science
Foundation of China (Grants No. 11275235, No. 11035006)
and the Chinese Academy of Sciences (the Knowledge Innovation
Project under Grant No. KJCX2-EW-N01).

\end{document}